\documentclass[twocolumn,epjc3]{svjour3}  
\smartqed  
\RequirePackage{graphicx}
\RequirePackage{color}
 \RequirePackage{mathptmx}      
\RequirePackage{latexsym}
\RequirePackage[numbers,sort&compress]{natbib}
\RequirePackage[colorlinks,citecolor=blue,urlcolor=blue,linkcolor=blue]{hyperref}
\RequirePackage{hyperref}
\RequirePackage{dsserif}
\RequirePackage[normalem]{ulem}

\journalname{Eur. Phys. J. Plus {\color{red}(Accepted for publication)}}
\begin{document}

\title{Supersolid-like square- and triangular-lattice crystallization of dipolar droplets  in a box trap
}

\titlerunning{Supersolid-like square- and triangular-lattice crystallization of dipolar droplets}        

\author{Luis E. Young-S.\footnote{e1,addr1}
        \and
        S. K. Adhikari\footnote{e2,addr2} 
}



\institute{$^*$Grupo de Modelado Computacional, Facultad de Ciencias Exactas y Naturales, Universidad de Cartagena, 
130014 Cartagena, Bolivar, Colombia \label{addr1}
           \and
           $^{**}$Instituto de F\'{\i}sica Te\'orica, UNESP - Universidade Estadual Paulista, 01.140-070 S\~ao Paulo, S\~ao Paulo, Brazil \label{addr2}
}

\date{Received: date / Accepted: date}

\maketitle

\begin{abstract}
 Using a beyond-mean-field model including a  Lee-Huang-Yang-type interaction,
we demonstrate a supersolid-like spatially-periodic square- and triangular-lattice  crystallization of droplets  
in a polarized dipolar condensate confined by an appropriate  three-dimensional (3D) box trap. 
In this paper
we consider a  rectangular box (cuboid) trap,  a square box (cuboid with two equal sides)  trap,    a  cylindrical box trap and a hexagonal box (hexagonal prism) trap. 
The droplet lattice is always formed in the   $x$-$y$ plane  perpendicular to the polarization $z$ direction of dipolar atoms. In contrast to a harmonic trap,
the box traps allow the formation of a large clean supersolid-like
 spatially-periodic crystallization in free space
without any distortion.   
Moreover, a droplet lattice can be formed in a 3D box trap with a significantly reduced number of atoms than in a harmonic trap, which could facilitate the experimental observation of  droplet lattice in a box trap.   
With present know-how such a supersolid-like crystallization of dipolar droplets in a 3D box trap  can be realized in a laboratory thus allowing the study of a large periodic lattice of dipolar droplets in free space bounded by rigid walls.
\end{abstract}

\section{Introduction}
\label{intro}
A supersolid \cite{sprsld,sprsld1,sprsld2,sprsld3,sprsld4,sprsld5} is a special quantum state of matter with  a rigid spatially-periodic  crystalline  structure \cite{rom}, breaking continuous translational invariance, that flows with zero viscosity as a superfluid  breaking  continuous gauge invariance.  Hence, a supersolid  simultaneously possesses the properties of a crystalline solid and a superfluid in contradiction with the belief that a frictionless flow is a property exclusive to a superfluid. 
The initial search
of  supersolidity in solid  $^4$He \cite{4} was not successful \cite{5}. However, there had been theoretical suggestions for creating a supersolid in a   spin-orbit (SO) coupled spinor Bose-Einstein condensate (BEC) \cite{7c,7C,7D},
in a  dipolar BEC \cite{7a,santos},  {and in a BEC with finite-range atomic interaction \cite{xxx}}.
After the recent experimental  observation of supersolids in a  quasi-one-dimensional (quasi-1D) SO-coupled pseudo spin-1/2 spinor BEC \cite{18} 
and in a  quasi-1D  \cite{1d2,1d3,1d4}  polarized dipolar BEC, 
the study of supersolids has  drawn a great deal of attention among   research workers  in low-temperature and condensed matter physics.  Specifically, for an appropriate number of atoms $N$ and for an appropriate mixture of contact and dipolar interactions,   
high-density droplet-formation was  observed experimentally  in a  strongly dipolar harmonically-trapped BEC    of $^{164}$Dy \cite{drop1} and  $^{168}$Er \cite{drop2} atoms. 
{For a  strongly dipolar BEC and for a   sufficiently large number of atoms $N$,} as   the density of atoms reaches a critical value, due to the  strong dipolar attraction, the trapped dipolar BEC  shrinks to a very small size   and a tiny droplet of size much smaller than the harmonical oscillator trap lengths,  is formed  as first observed \cite{2d3} in a  quasi-two-dimensional (quasi-2D) dipolar BEC of 
$^{164}$Dy atoms. However, 
such a droplet can accommodate  a maximum number of atoms.
With further increase of $N$, multiple droplets arranged along a straight line \cite{1d2,1d3,1d4,1d1,1d5}, 
for a quasi-1D configuration,  
or 
on  a spatially-periodic  triangular  \cite{2d3,2d2} or square \cite{ly} lattice, in a quasi-2D configuration, 
can be  formed.

In the framework of a mean-field  Gross-Pitaevskii (GP) equation,
a  sufficiently strongly  dipolar BEC, for a large number of atoms or  for  a large atomic dipole moment,  collapses. Nevertheless,  
 a beyond-mean-field Lee-Huang-Yang-type \cite{lhy} (LHY-type)   repulsive   interaction \cite{qf1,qf2}, with a higher-order quartic  repulsive nonlinearity, compared to the cubic nonlinearity of the GP equation, 
can stop the  collapse and 
form  a   stable droplet in a  harmonically-trapped strongly dipolar  BEC \cite{santos,drop3,blakie}. This beyond-mean-field model with the LHY-type interaction \cite{qf1,qf2}
has been successfully used in the study of droplets and droplet lattice in a harmonically-trapped dipolar BEC in both quasi-1D \cite{1d2,1d3,1d4,1d1,1d5}  and quasi-2D \cite{2d3,2d2,2d1,2d4} configurations.

In a quasi-1D harmonically-trapped dipolar BEC  of polarized 
$^{164}$Dy \cite{1d4,1d1,1d5},  $^{162}$Dy \cite{1d2,1d3,1d7}, and $^{166}$Er \cite{1d4,1d5}  atoms, a spontaneous spatially-periodic crystallization of droplets in a straight line was observed, whereas in a quasi-2D
harmonically-trapped dipolar BEC  of polarized 
$^{164}$Dy atoms \cite{2d3,2d2}, a spatially-periodic  crystallization of droplets on
a triangular lattice was observed.  These linear \cite{1d6,1d8}
and triangular-lattice \cite{2d4,blakieprl,other1,other2,other3}  crystallization of droplets were also 
confirmed   in  theoretical studies using three-dimensional (3D) beyond-mean-field models with the LHY-type interaction. More recently, the formation of a spatially-periodic square-lattice pattern  of droplets in a harmonically-trapped dipolar BEC  was demonstrated \cite{ly} in a theoretical study using a beyond-mean-field model.
{Nevertheless, in most of these investigations, specially in the
numerical studies on a truncated finite system as in this paper, the  relation of these quasi-1D and quasi-2D spatially-periodic 
states  
to supersolidity was not  established \cite{44}. That task requires a demonstration of 
the spontaneous breakdown
of gauge symmetry (giving the superfluid order parameter)
and the spontaneous breakdown of translational
symmetry   in these systems. 
Lacking such a rigorous demonstration of supersolidity,  in this paper we call  these spatially-periodic
states supersolid-like states as in some previous 
studies on quasi-2D SO-coupled spinor BECs \cite{17,19}.}

In this paper, we demonstrate and study numerically by imaginary-time propagation  a spatially-periodic quasi-2D  crystallization of dipolar droplets  
on  triangular and square lattices through the formation of droplet-lattice states 
in the $x$-$y$ plane,
 perpendicular to the polarization $z$ direction of dipolar atoms,  
in a 3D box trap, using a beyond-mean-field model including the LHY-type interaction.
{The dynamical stability of these droplet-lattice states was established by real-time propagation over a long period of time after introducing a perturbation at time $t=0$.}
One has a uniform potential inside a box trap and an infinite potential outside,  
  such that the wave function is zero on the boundary of the box.   Thus a 3D box trap allows 
the formation and study of droplets in free space with a uniform 
potential bounded by a rigid wall, without an interfering harmonic oscillator potential, as in other studies, thus resulting in a few advantages. 
We find that 
an increase in the size of the box trap allows  the formation of  a very large lattice of droplets in a uniform potential  with a significantly smaller number of atoms per droplet without any visible distortion, and with a  much reduced  uniform  background atom cloud, 
when compared to the {previous results \cite{ly,2d4} obtained with} a  harmonic trap,  which could be an advantage for the  experimental   study of a spatially-periodic quasi-2D  crystallization of dipolar droplets in a box trap.

Recently,  BECs were created experimentally in a quasi-1D \cite{16a,16a1,16b},   and 3D \cite{16c,17a,17b,17c,17d} box traps. The quasi-1D \cite{16a,16a1,16b} trap has uniform  confinement along one direction and harmonic confinement in transverse directions. The 3D optical trap of Refs. \cite{16c,17a,17b,17c,17d} is a uniform 3D  cylindrical box trap which has a  circular box trap in  $x$-$y$ plane
with  a uniform confinement in    $z$ direction.  The experimental technique now allows the creation 
of a combination of box trap and harmonic trap in $x,y$ and $z$  directions.
In  the   $x$-$y$ plane one can create a circular box trap or a square box trap coupled to   a box trap along the   $z$ direction.  Box traps are now regularly used in experimental
and theoretical studies of cold atoms \cite{bt1,bt2,bt3,bt4,bt5}. 
In this paper we will consider the following traps:  a 3D square  box (a cuboid with two equal sides)  trap $U_A$,    a 3D rectangular box (a cuboid)
 trap $U_B$,  a cylindrical box trap  $U_C$, and  a hexagonal box  (hexagonal prism) trap $U_D$. We demonstrate the formation of  a square-lattice crystallization of droplets in traps $U_A$, $U_B$ and $U_C$ and a triangular-lattice crystallization of droplets in traps $U_B$, $U_C$ and $U_D$. 
 We find that the symmetry of the final state in numerical simulation is sensitive to the initial state employed.  A final state with a specific symmetry $-$ a square, or a triangular lattice $-$ can easily be obtained with the use of  an initial state with the same symmetry. 
 No such supersolid-like state can be obtained in a trapped BEC with isotropic contact interaction.  In a dipolar BEC, a single droplet is  stable for the number of atoms below a critial value $N_{\mathrm{crit}}$. {As the number of atoms increases beyond $N_{\mathrm{crit}}$, a single-droplet state becomes energetically unstable and collapses and stable 
 multiple droplets are created  and due to an  interplay between the
 binding box potential and }
 dipolar repulsion in the $x-y$ plane, a supersolid-like arrangement of droplets is formed.

In Sec. \ref{II} the  beyond-mean-field model with the  LHY-type interaction is presented.  
In Sec. \ref{III} we present the numerical results for stationary  states with  two types of periodic array of droplets, e.g. square and  triangular lattice,  in a   dipolar BEC confined by an appropriate box trap.  
A square-lattice arrangement of droplets is possible in a square box trap $U_A, U_B$ and $U_C$. For the formation of a triangular-lattice arrangement of droplets we considered an appropriate rectangular box trap $U_B$, a cylindrical box trap $U_C$, and a hexagonal box trap $U_D$. The square box trap $U_A$  is the most appropriate for generating a square lattice  of dipolar droplets  and the hexagonal box trap $U_D$ is the most suitable for generating a triangular lattice of droplets. 
    Finally, in Sec. \ref{IV} we present a summary of our findings.

\section{Beyond-Mean-field model}

\label{II}

In this study we consider a 3D beyond-mean-field model   including an appropriate    LHY-type interaction \cite{qf1,qf2}. 
We consider a  BEC of $N$  dipolar  atoms,  of mass $m$ each, polarized along the $z$ direction,
interacting through the following 
atomic contact and dipolar interactions   \cite{dipbec,dip,yuka}
\begin{eqnarray}
V({\bf R})= \frac{4\pi \hbar^2 a}{m}\delta({\bf R })+
\frac{\mu_0 \mu^2}{4\pi}\frac{1-3\cos^2 \theta}{|{\bf R}|^3}
,
\label{eq.con_dipInter} 
\end{eqnarray} 
where $\mu_0$ is the permeability of vacuum, $a$ is the scattering length,   $\mu$ is the magnetic moment of each atom,
${\bf R}= {\bf r} -{\bf r}'$ is the vector joining two atoms placed at $\bf r \equiv \{x,y,z\}$ and $\bf r' \equiv \{x',y',z'\}$
with $\theta$  the angle between vector   ${\bf R}$ and the  
$z$ axis. 
It is convenient to introduce the following  dipolar length to quantify 
the strength of dipolar 
interaction  
\begin{eqnarray}
a_{\mathrm{dd}}=\frac{\mu_0 \mu^2 m }{ 12\pi \hbar ^2}.
 \label{eq.dl}
 \end{eqnarray}
The dimensionless ratio 
 \begin{eqnarray}
\varepsilon_{\mathrm{dd}}\equiv \frac{a_{\mathrm{dd}}}{a}
\end{eqnarray} 
defines
the relative strength of  dipolar and   contact interactions.

A dipolar BEC is described by the following  3D beyond-mean-field  equation with the   LHY-type interaction \cite{blakie,2d4,dipbec,dip,yuka}
\begin{eqnarray}\label{eq.GP3d}
{\mathrm i} \hbar \frac{\partial \psi({\bf r},t)}{\partial t} &=&\frac{\hbar^2}{m}
{\Big [}  -\frac{1}{2}\nabla^2
+\frac{m}{\hbar^2}U({\bf r})
+ {4\pi }{a} N \vert \psi({\bf r},t) \vert^2 \nonumber\\
& +& {3}a_{\mathrm{dd}}  N
\int \frac{1-3\cos^2 \theta}{|{\bf R}|^3}
\vert\psi({\mathbf r'},t)\vert^2 d{\mathbf r}'  
\nonumber 
 \\
&
 +&{\gamma_{\mathrm{QF}}}
|\psi({\mathbf r},t)|^3
\Big] \psi({\bf r},t), 
\end{eqnarray}
where ${\mathrm i}=\sqrt{-1},$ $U({\bf r})  $ is the box trap and
the wave-function normalization is $\int \vert \psi({\bf r},t) \vert^2 d{\bf r}=1$. The third and the fourth terms on the right-hand-side of  Eq. (\ref{eq.GP3d}) are contributions of atomic contact and dipolar interactions and the last term is the contribution of the beyond-mean-field   LHY-type interaction with the
 coefficient $\gamma_{\mathrm{QF}}$
 given by \cite{qf1,qf2,blakie}
\begin{eqnarray}\label{qf}
\gamma_{\mathrm{QF}}= \frac{128}{3}\sqrt{\pi a^5} Q_5(\varepsilon_{\mathrm{dd}}),
\end{eqnarray}
where the  function
\begin{equation}
 Q_5(\varepsilon_{\mathrm{dd}})=\ \int_0^1 dx(1-x+3x\varepsilon_{\mathrm{dd}})^{5/2} 
\end{equation}
 can be evaluated to yield the following expression for the  coefficient $\gamma_{\mathrm{QF}}$ \cite{blakie}
\begin{eqnarray}\label{exa}
\gamma_{\mathrm{QF}}&=&\ \frac{128}{3}\sqrt{\pi a^5} 
\frac{(3\varepsilon_{\mathrm{dd}})^{5/2}}{48}  \left[(8+26\epsilon+33\epsilon^2)\sqrt{1+\epsilon}\right.\nonumber\\
& + &\left.
\ 15\epsilon^3 \mathrm{ln} \left( \frac{1+\sqrt{1+\epsilon}}{\sqrt{\epsilon}}\right)  \right], \quad  \epsilon = \frac{1-\varepsilon_{\mathrm{dd}}}{3\varepsilon_{\mathrm{dd}}}.   
\end{eqnarray}
Actually, the function $Q_5(\varepsilon_{\mathrm{dd}}),$ as well as the coefficieq.ent  $\gamma_{\mathrm{QF}}$, representing a beyond-mean-field  LHY-type correction for dipolar atoms, is complex for
$ \varepsilon_{\mathrm{dd}} > 1$ and, for studies of stationary states, expression (\ref{exa}) is formally   meaningful  only for $ \varepsilon_{\mathrm{dd}} \le 1$, where this expression
is real \cite{27}.  However its imaginary part remains small compared to its real part for medium values of   $a$
used in this paper $a\approx  85a_0$   and will be neglected in this study of stationary self-bound states as in other studies.

Equation (\ref{eq.GP3d}) can be reduced to 
the following  dimensionless form by scaling lengths in units of  a length scale $l$, time in unit of $ml^2/\hbar$,  energy in unit of $\hbar^2/ml^2 $, and density $|\psi|^2$ in unit of $l^{-3}$
\begin{eqnarray}
{\mathrm i} \frac{\partial \psi({\bf r},t)}{\partial t} &=&
{\Big [}  -\frac{1}{2}\nabla^2
+ U({\bf r}) + 4\pi{a} N \vert \psi({\bf r},t) \vert^2
\nonumber\\ &
+&3a_{\mathrm{dd}}  N
\int \frac{1-3\cos^2 \theta}{|{\bf R}|^3}
\vert\psi({\mathbf r'},t)\vert^2 d{\mathbf r}'  \nonumber \\ &
+& \gamma_{\mathrm{QF}}N^{3/2}
|\psi({\mathbf r},t)|^3
\Big] \psi({\bf r},t),
\label{GP3d}
\end{eqnarray} 
where we are using the same functions to represent the scaled (and original unscaled) variables.

 In this study we  will consider the following box traps: ($U_A$) a  square box (a cuboid with two equal sides) trap defined by
 $U({\bf r}) =0$ for $|x| < l_1$, $|y| < l_2=l_1$ and $|z| < l_3$ and infinite otherwise, where $l_1,l_2, l_3$ are the lengths of the orthogonal box trap;
 ($U_B$) a rectangular  box (a cuboid) trap defined by 
 $U({\bf r}) =0$ for $|x| < l_1$, $|y| < l_2$ and $|z| < l_3$ and infinite otherwise;
  ($U_C$) a 3D cylindrical box trap, now routinely used in experiments \cite{16c,17a,17b}, defined by   
  $U({\bf r}) =0$ for $\rho \equiv \sqrt{x^2+ y^2} < l_1$ and $|z| < l_3$ and infinite otherwise, where $l_1$ is the radius of the cylindrical box trap and $l_3$ is its length;
 ($U_D$) a 3D hexagonal (hexagonal prism) box trap defined by   
  $U({\bf r}) =0$ for $|x| <\sqrt {3}l_1/2 $, \, $|\pm x+\sqrt{3}y| < \sqrt{3}l_1$ and    $|z| < l_3$ and infinite otherwise, where $l_1$ is the length of each side of the hexagonal box and $l_3$ is its length.

\section{Numerical Results}

\label{III}

The   partial differential equation (\ref{GP3d}) for a dipolar BEC is solved   numerically employing  the split-time-step Crank-Nicolson
method \cite{crank} by time propagation, using 
C or FORTRAN programs \cite{dip} or their open-multiprocessing versions \cite{omp}.  { We used imaginary-time propagation for a stationary state and real-time propagation for studying dynamics \cite{crank,crank1}.  }
Often, the density of the system has large extension in the $x$-$y$ plane and it is appropriate to take a smaller number of discretization steps along $z$ direction as compared to the same along $x$ and $y$ directions.  {We take space step $\Delta = 0.05$ in all three directions $-$ $x,y,$ and $z$ $-$ 
and a time step $0.1\Delta^2$ in imaginary-time propagation and a time step $0.05\Delta^2$ in real-time propagation
 and verified that the presented numerical results  converged with respect to a reduction of space ane time steps.  }
Due to the divergent $1/|{\bf R}|^3$ term,
it is problematic to treat  the nonlocal dipolar interaction integral numerically in the beyond-mean-free model (\ref{GP3d}) in configuration space. To avoid this problem, the dipolar term is calculated in the momentum space by a Fourier transformation using a convolution identity \cite{dip}, which is economic numerically due to the smooth behavior of this term in momentum space. The Fourier transformation of the dipolar potential has an analytic result \cite{dip}
 which enhances  the accuracy of the numerical procedure. After obtaining the solution in momentum space, the result in configuration space is obtained by an inverse Fourier transformation.

We present results for a BEC of strongly dipolar $^{164}$Dy atom, as many of the experimental \cite{drop1,2d3,2d2} and related theoretical   \cite{2d4,blakieprl,other1,other2,other3}  studies  were performed with this atom.
In this study we will take the scaling length $l=1$ $\mu$m as the unit of length, so that the unit of density 
$|\psi|^2$ is 1 $\mu$m$^{-3}$ and evaluate the results in dimensionless units.
  For the formation of a square- or a triangular-lattice of droplets we need a strongly dipolar atom with $\varepsilon_{\mathrm{dd}}>1$   \cite{2d3}, while the dipolar interaction dominates over the contact interaction. 
The experimental estimate of scattering length of $^{164}$Dy atom is  $a=(92\pm 8)a_0$  \cite{scatmes}
and  its dipolar length is $a_{\mathrm{dd}}=130.8a_0$ corresponding to the magnetic dipole moment $10\mu_{\mathrm{B}}$, where $a_0$  is the Bohr radius and $\mu_{\mathrm{B}}$ is the Bohr magneton.
In this study we take $a=85a_0$, as in a previous study of droplets \cite{ly}; consequently, $\varepsilon_{\mathrm{dd}}=1.5388... >1$.
This value of scattering length was used in a previous study of droplet lattice in presence of harmonic trapping potentials \cite{ly} and 
is close to its experimental estimate and also to  the values  $a=88a_0$ \cite{2d2,2d1,2d4} and $a=70a_0$ \cite{blakieprl}
used in other   studies of formation of  quantum droplets in a   quasi-2D harmonically trapped dipolar BEC.

An appropriate choice of the initial state is essential for an efficient evaluation  of a droplet-lattice arrangement  employing imaginary-time propagation \cite{ly}. The numerical simulation of a few-droplet state can be performed by employing a Gaussian initial function, viz. Fig. \ref{fig1}.
 The numerical evaluation of a large square- or a triangular-lattice  crystallization was started by an initial analytic  wave function  with a lattice structure of appropriately placed Gaussian droplets on a square or a triangular lattice, respectively \cite{ly}.   For example, the numerical simulation of a 81-droplet  $9 \times 9$ square-lattice state was initiated by the following  analytic initial wave function with 81 Gaussian droplets appropriately placed on a square  lattice:
\begin{eqnarray}\label{ana1}
\phi({\bf r})\sim \sum_{i,j=0}^{\pm 1, \pm 2,\pm 3, \pm 4}e^{-(x+i\beta)^2-(y+j\beta)^2}e^{-z^2/\alpha^2}
\end{eqnarray}
with the lattice constant $\beta \approx 5$ and an appropriately chosen $z$-width $\alpha$. Similarly, the numerical simulation of a 37-droplet triangular-lattice state is initiated by the following analytic initial function with 37 droplets appropriately placed on a triangular lattice:
\begin{eqnarray}\label{ana2}
\phi({\bf r})&\sim&  \left[\sum_{i=0}^{\pm 1, \pm 2,\pm 3}e^{-(x+4i\beta)^2-y^2}  \right. \nonumber \\&+&
\sum_{i=\pm 1, \pm 3, \pm 5}\sum_{j=\pm 1}  e^{-(x+2i\beta)^2-(y+2\sqrt 3 j\beta)^2}  \nonumber \\&+&
\sum_{i=0, \pm 2, \pm 4}
\sum_{j=\pm 2}e^{-(x+2i\beta)^2-(y+2\sqrt 3 j\beta)^2}  \nonumber \\&+&
\left.
\sum_{i=\pm 1, \pm 3}
\sum_{j=\pm 3}e^{-(x+2i\beta)^2-(y+2\sqrt 3 j\beta)^2}  \right]e^{-z^2/\alpha^2}.
\end{eqnarray}
The initial analytic functions in other cases are chosen appropriately.

 { 

\begin{figure}[t!]
\begin{center} 

 \includegraphics[width=\linewidth]{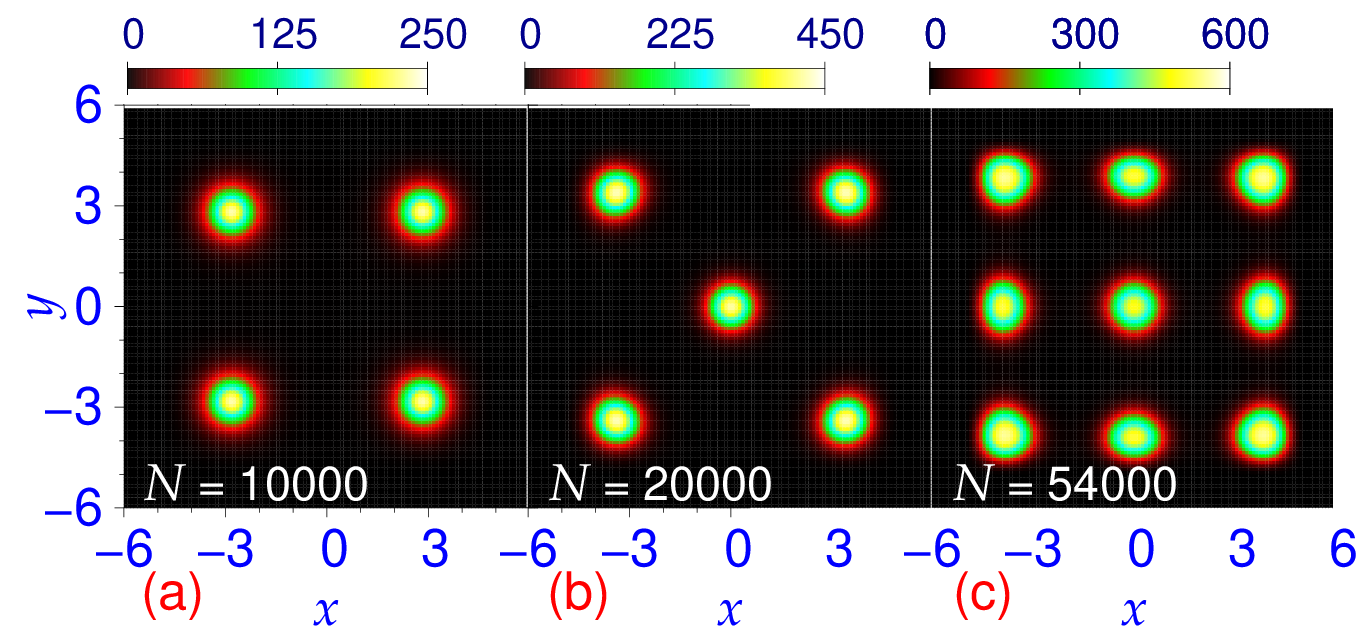}

\caption{ Contour plot of  density of  $^{164}$Dy atoms  $N|\psi(x,y,0)|^2$ in the $x$-$y$ plane    in (a) square-lattice crystallization   of   4 droplets,   (b)  centered square-lattice crystallization of     5 droplets,  and (c) square-lattice crystallization   of   9 droplets,
  in a 3D cube box trap $U_A$. The initial state used in the evaluation of these states  was a {\it single Gaussian} state which evolves to multiple-droplet final states. 
The $x,y$ and $z$ lengths of the box potential are  $l_1=l_2=l_3=5$ and  $a=85a_0$.
 Plotted quantities in all figures are dimensionless and the corresponding  number of atoms $N$ is displayed in all plots of this paper. The actual 3D density can be obtained using the length scale $l=0.607$ $\mu$m; for example, the maximum density in (c) is $604/l^3\approx 2.8 \times 10^{21}$ m$^{-3}$. }
\label{fig1} 
\end{center}
\end{figure}

For a harmonically-trapped specific dipolar BEC with a given $a$ and $a_{\mathrm{dd}}$ 
there is a typical lattice constant (length).
For  a dipolar BEC of  $^{164}$Dy atoms  with $a=85a_0$ this lattice constant is approximately $\sim 4$ $\mu$m in the trap of Ref.    \cite{ly}. As the number of atoms in that  harmonically-trapped  $^{164}$Dy BEC is increased,  more and more droplets are formed and a larger lattice is generated filling a larger space maintaining approximately  the same lattice spacing. This lattice spacing for a dipolar BEC of  $^{164}$Dy atoms is approximately the same for a  box trap. {
A square-lattice state is efficiently generated in a  square or a rectangular box trap, whereas a triangular-lattice state is  efficiently generated in a hexagonal box trap. Because of the rigid walls of the box trap and because of large long-range dipolar repulsion the droplets near the boundary will be arranged with the same symmetry as the box potential which will influence the droplets further in the bulk to follow the same symmetry and eventually fill in the whole inner space. }
To generate 
a specific droplet-lattice state one should consider a  box trap which could accommodate the same. 
 For example, as the lattice constant is approximately $\sim 4 $  in the box trap, the extension  of a 25-droplet $5\times 5$ square-lattice state in the 
$x$-$y$ plane   is about $20\times 20$. Hence,  the $x$ and $y$ lengths of a square  box trap $U_A$ should be about $l_1=l_2\approx 10$.  For a 49-droplet square-lattice state the $x$ and $y$ lengths should be slightly larger and about $l_1=l_2\approx 15$.
The extension of a droplet in the $z$ direction is about a few     length units   and in all calculations we will take the length of the  box trap to be $l_3= 5$. In addition
one should have an appropriate number of atoms in the BEC to form a certain number of droplets, as a minimum number of atoms is needed to form a new droplets.

\begin{figure}[t!]
\begin{center}
\includegraphics[width=\linewidth]{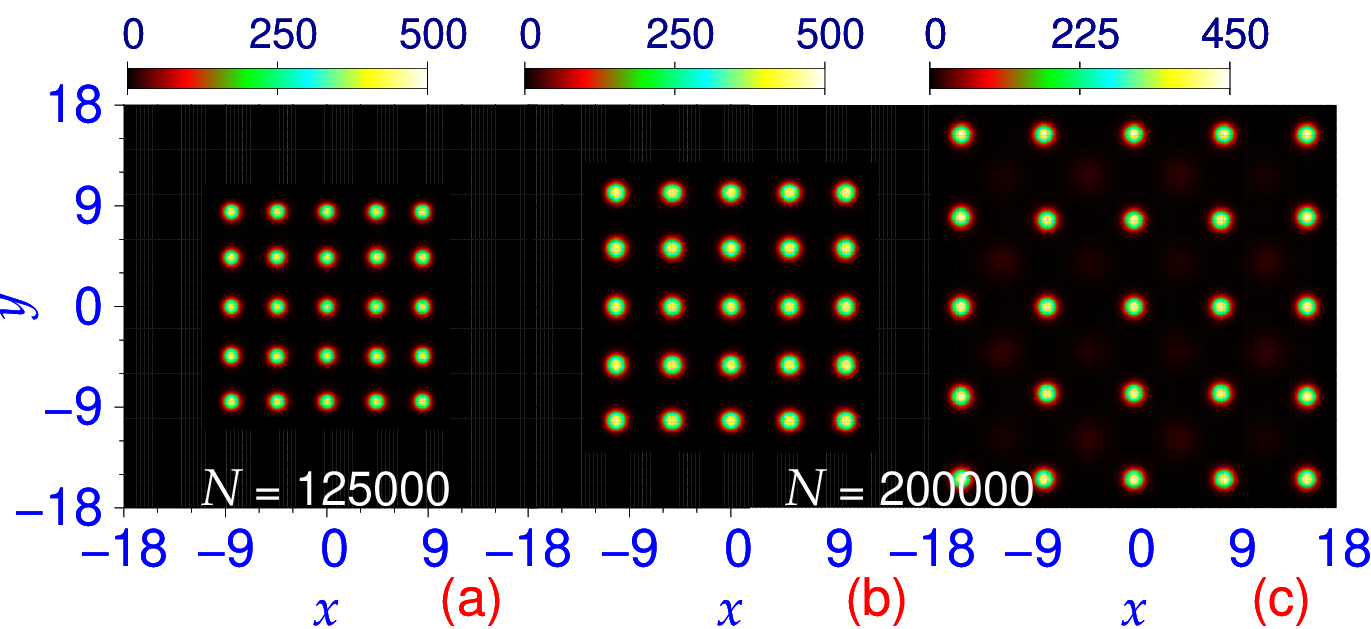} 
\includegraphics[width=\linewidth]{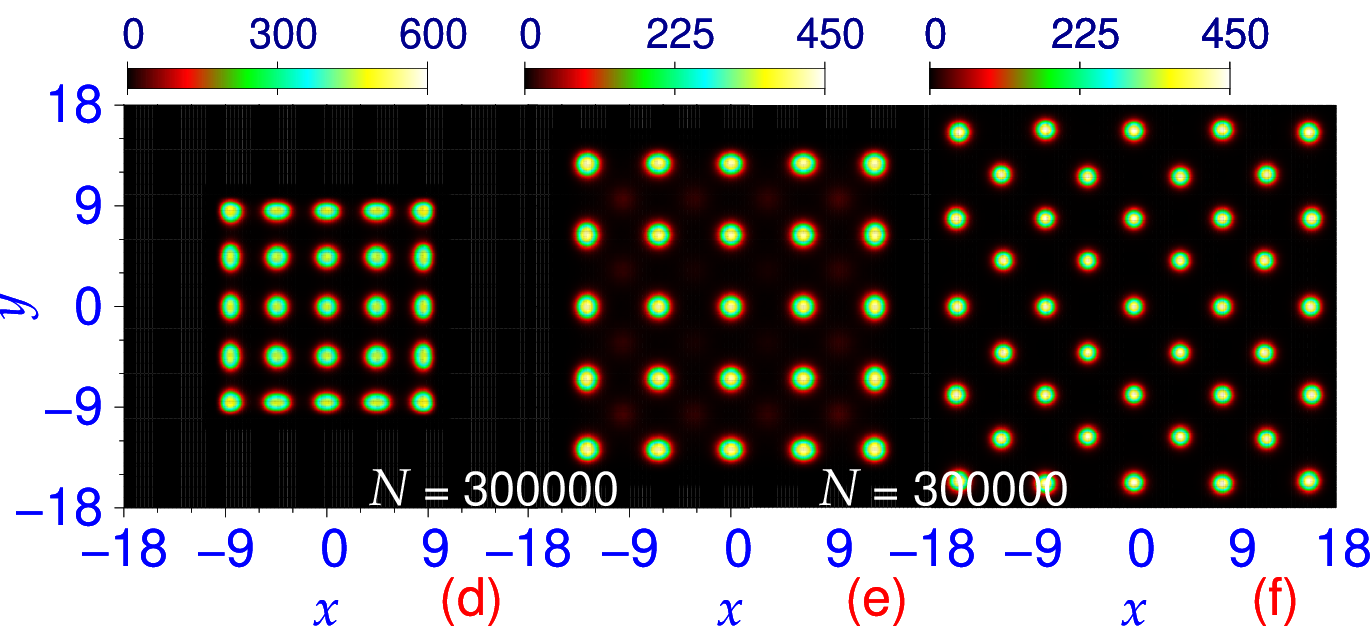}
 
\caption{  Contour plot of  density of  $^{164}$Dy atoms  $N|\psi(x,y,0)|^2$ in the $x$-$y$ plane  of square-lattice crystallization of 25 droplets 
for (a)  $N=125000$, with $x$ and $y$ lengths $l_1=l_2=10$,   (b)   $N=200000$, with   $l_1=l_2=12$,   (c)   $N=200000$, with  $l_1=l_2=18$,
 (d)   $N=300000$, with   $l_1=l_2=10$,
 (e)   $N=300000$, with   $l_1=l_2=15$,
(f)    $N=300000$, with   $l_1=l_2=18$, in a 3D square  box trap $U_A$. The $z$-length in all cases is $l_3=5$ and $a=85a_0$. The initial state in these imaginary-time simulations [including the state in (f)] is an analytic 25-droplet $5\times 5$ state  arranged on a square lattice, viz. Eq.  (\ref{ana1}).
}
\label{fig2} 
\end{center}
\end{figure}

First we consider the spontaneous formation of droplets on a square lattice  in the box trap $U_A$. For this purpose we consider a cubic box trap, which is a special case of a square box trap with equal $x$, $y$ and
 $z$-lengths: $l_1=l_2=l_3=5$. In numerical simulation by imaginary-time propagation,   we consider an analytic  Gaussian initial state with a small number of atoms ($N	< 5000$). The final 
state is also found to be an extended state occupying almost the whole box (result not shown here).   
As the number of atoms is increased  approximately to $N\approx 10000$, four droplets arranged on a square lattice are formed spontaneously from an analytic initial Gaussian state in imaginary-time calculation.
This is illustrated through a 
 contour plot of  density of  $^{164}$Dy atoms  $N|\psi(x,y,0)|^2$ in the $x$-$y$ plane in Fig. \ref{fig1}(a) for $N=10000$. 
We could not find a one-, two-, or three-droplet state in this cubic box trap $U_A$   for smaller $N$ ($N<10000$).   As the number of atoms is increased further to $N\approx 20000,$ five droplets arranged on a centered square lattice are formed as shown in Fig. \ref{fig1}(b)  through a contour plot of density 
of  $^{164}$Dy atoms  
$N|\psi(x,y,0)|^2$ in the $x$-$y$ plane 
for 
$N=20000$ atoms. The five-droplet state continue to exist as the unique possible droplet state for $N$ up to $ \approx 50000$, although the size of the droplet becomes larger as the number of atoms is increased. 
For larger values of $N$ ($N>  50000$),    for this cubic  box trap,
 a nine-droplet state appears in imaginary-time simulation from an analytic  Gaussian initial state as illustrated in Fig. \ref{fig1}(c)  through a contour plot of density $N|\psi(x,y,0)|^2$  of  $^{164}$Dy atoms  $N|\psi(x,y,0)|^2$ in the $x$-$y$ plane
  for 
$N=54000$ atoms.
We could not find a six-, seven-, or eight-droplet state in this cubic box trap $U_A$  for any  $N$, 
indicating that the system prefers these spatially-periodic ordered square-lattice  states in a square box trap.
We will see that these two arrangements $-$ square lattice  in (a) and (c)  and centered square lattice  in (b) $-$ are the most likely unit cells for large droplet-lattice formation in a square box trap 
as the size of the square box trap and number of atoms are increased so as to form   and accommodate  a large number of droplets on a square lattice 
for a large number of atoms in a larger space. A distorted lattice state will be generated  unless the size of the box trap is  increased as the number of atoms is increased.

\begin{figure}[t!]
\begin{center}
\includegraphics[width=.32\linewidth]{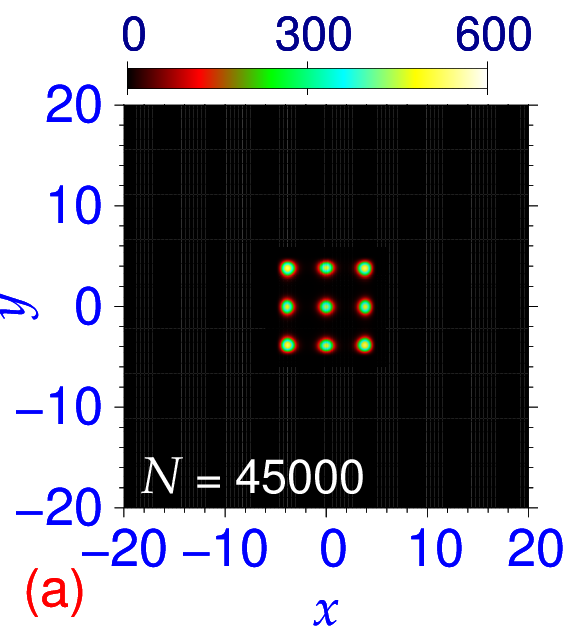}
\includegraphics[width=.32\linewidth]{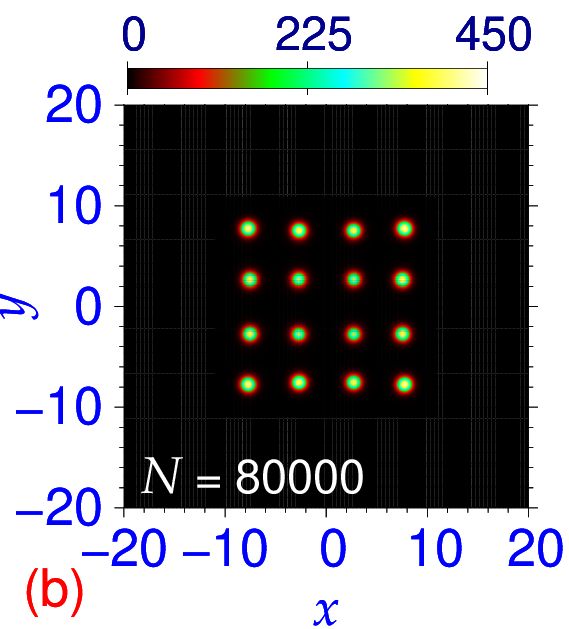}
\includegraphics[width=.32\linewidth]{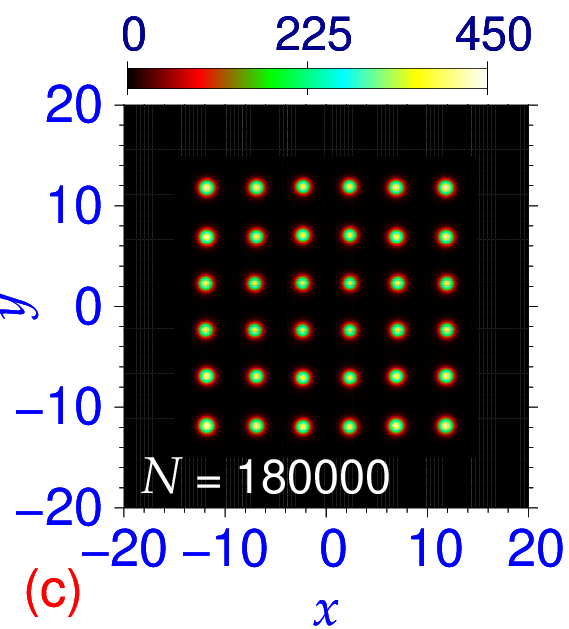}
\includegraphics[width=.32\linewidth]{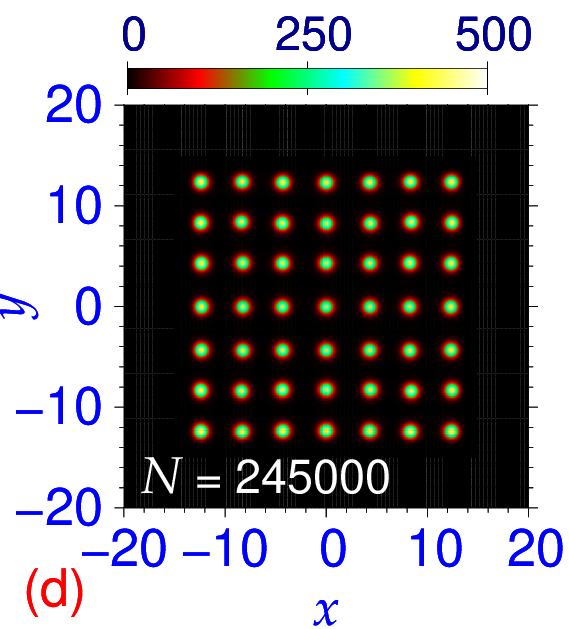} 
\includegraphics[width=.32\linewidth]{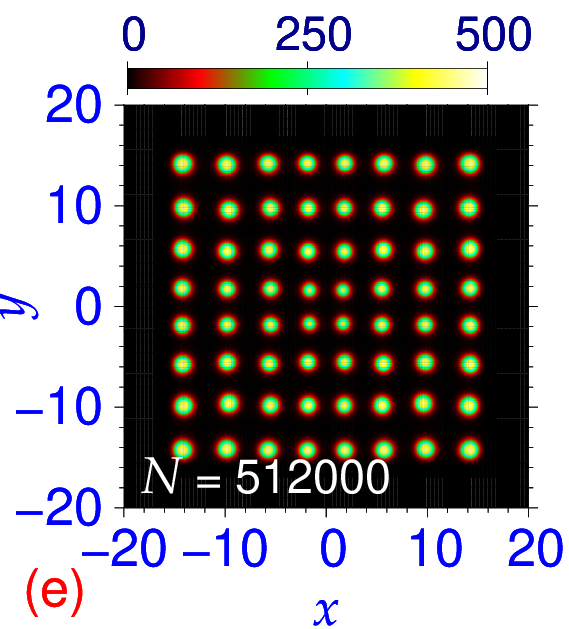}
\includegraphics[width=.32\linewidth]{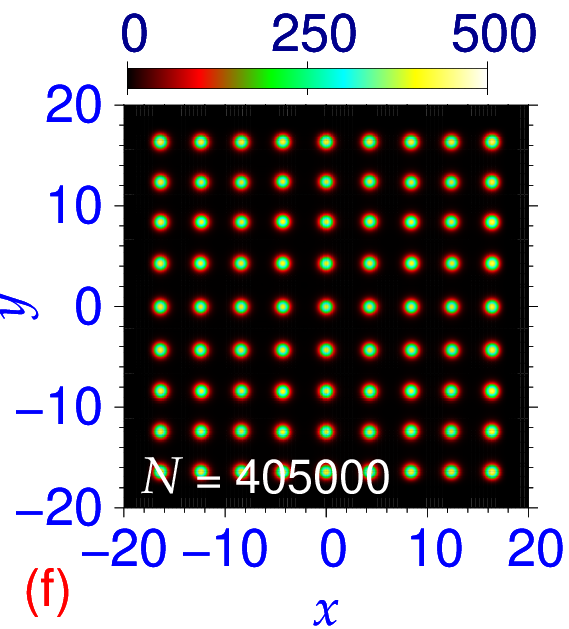}

\caption{ Contour plot of  density of  $^{164}$Dy atoms  $N|\psi(x,y,0)|^2$ in the $x$-$y$ plane  of square-lattice crystallization 
of (a)  9 droplets  with $x$ and $y$ lengths $l_1=l_2=5$,   (b)   16 droplets with  $l_1=l_2=10$,
(c)  36 droplets with   $l_1=l_2=15$,   (d) 49 droplets   with   $l_1=l_2=15$,
(e)  64 droplets with   $l_1=l_2=16$,   (f) 81 droplets   with   $l_1=l_2=18$,
 in a 3D square box trap $U_A$. The $z$-length in all cases is $l_3=5$ and $a=85a_0$. 
}
\label{fig3} 
\end{center}
\end{figure}

To demonstrate the formation of a specific droplet-lattice state for different number of atoms in the square box trap $U_A$,    we consider  a 25-droplet square-lattice state for different number of atoms and different sizes (lengths $l_1=l_2$) of a square box trap $U_A$. In Fig. \ref{fig2} we display  the
 contour plot of  density of  $^{164}$Dy atoms  $N|\psi(x,y,0)|^2$ in the $x$-$y$ plane 
 for (a) $N=125000$, $l_1=l_2=10$,    (b) $N=200000$, $l_1=l_2=12$,  (c) $N=200000$, $l_1=l_2=18$, 
 (d) $N=300000,$ $l_1=l_2=10$, (e) $N=300000$, $l_1=l_2=15$, 
 (d) $N=300000,$ $l_1=l_2=18$. In all cases the $z$-length is $l_3=5$ and 
the initial state in imaginary-time propagation 
was an analytic  25-droplet  state arranged on a $5\times 5$  square lattice, viz. Eq.  (\ref{ana1}). 
 In the first five cases displayed in (a)-(e), the final state is also a  25-droplet state arranged on a $5\times 5$  square lattice,  although the size of the droplet lattice (and also that of the unit cell) increased for larger $x$ and $y$ lengths. The scenario remains the same with the increase of the number of   atoms  $N$ to about $\approx 240000$, beyond which  the number of droplets for large $x$ and $y$ lengths ($l_1=l_2\ge 18$) spontaneously
increases from 25 (in the initial state)
to 41 (in the final state) during imaginary-time propagation. 
If we compare plots (e) and (f), we  find that with the increase in the size of the box trap from $l_1=l_2=15 $ to $18$ for  the fixed number $N=300000$ of atoms,   the number of droplets has spontaneously
increased from 25 
to 41  in the final state  
  arranged in a square lattice inclined at an angle of 45$^{\circ}$ with respect to the initial square lattice. 
This 41-droplet state is composed of alternately placed five rows of five-droplet states and four rows of four-droplet states. 
 The  41-droplet   square-lattice state can be formed for $350000 > N > 240000$ and  for large values of  $x$ and $y$ lengths $l_1, l_2$.
For smaller values of $l_1$ and $l_2$ the number of droplets remain 25 in the final state and for large $N$ ($N > 350000$), a larger number of droplets ($N>41$) arranged on a distorted lattice without a periodicity can be formed (result not presented here).
In plots (a)-(e) the unit cell of the lattice arrangement 
is the square lattice of Fig. \ref{fig1}(a) and the same in plot (f) is the centered square lattice of Fig. \ref{fig1}(b).  
All the states displayed in Fig. \ref{fig2} have a maximum density at the center $x=y=0$ and are 
parity-symmetric in $x$ and $y$.

\begin{figure*}[t!]
\begin{center}
 
\includegraphics[width=\textwidth]{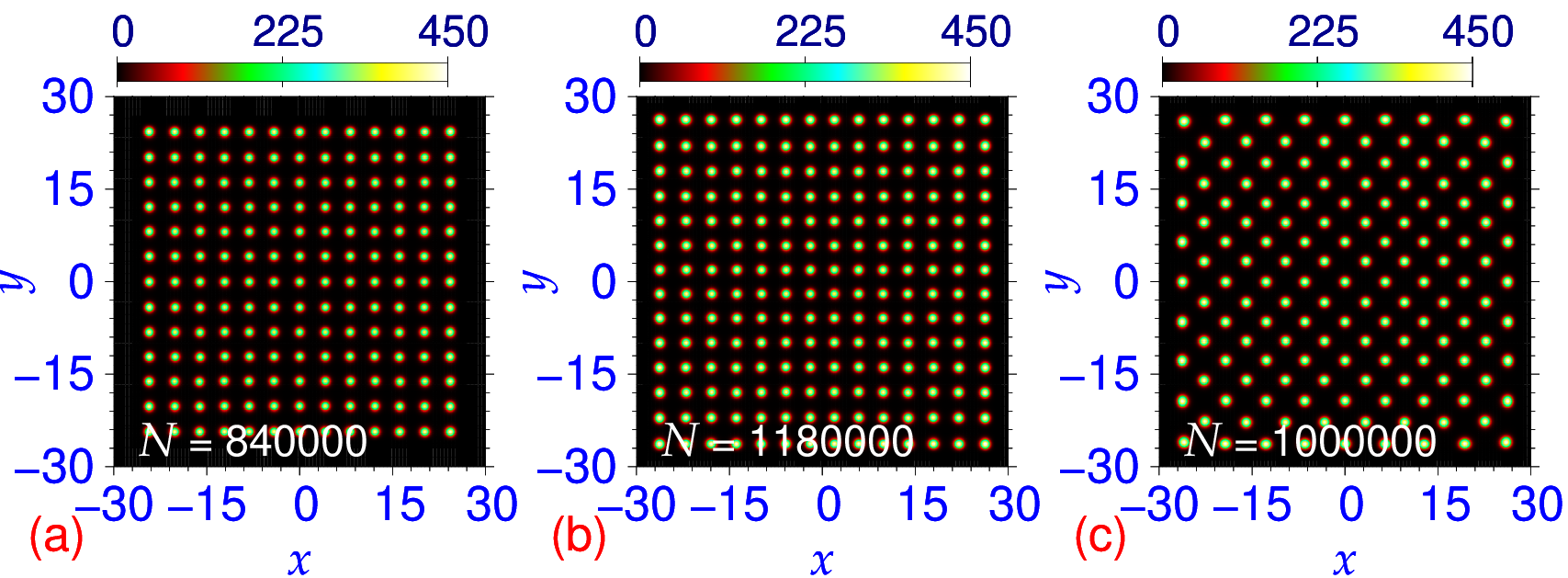}

\caption{  Contour plot of  density of  $^{164}$Dy atoms  $N|\psi(x,y,0)|^2$ in the $x$-$y$ plane   of square-lattice crystallization 
in a (a)  169-droplet $13\times 13$ square-lattice state  with $N=840000$ and  $x$ and $y$ lengths $l_1=l_2=26$,   (b)   196-droplet $14\times 14$ square-lattice state with $N=1180000$ and   $l_1=l_2=28$,
and   (c)   145-droplet state  with  $N=1000000$ and  $l_1=l_2=28$,
 in a 3D square box trap $U_A$. The unit cell in (a) and (b) is the square lattice of Fig. \ref{fig1}(a) and that in (c) is the centered square lattice of Fig. \ref{fig1}(b).
The $z$-length in all cases is $l_3=5$ and $a=85a_0$.  }

\label{fig5} 
\end{center}
 
\end{figure*}

After having established the formation of a 25-droplet $5\times 5$ square-lattice state in a dipolar BEC of $^{164}$Dy atoms, 
we study the formation of 9-, 16-, 36-, 49-, 64-, and  81-droplet
$3\times 3$, $4\times 4$, $6\times 6$, $7\times 7$, $8\times 8$, and $9\times  9$ 
 square-lattice states in a square box trap $U_A$
and the results are displayed, respectively,  in Fig. \ref{fig3} for (a)  $N=45000$, $x$ and $y$ lengths $l_1=l_2=5$, 
(b)  $N=80000$,   $l_1=l_2=10$, (c)  $N=180000$,  $l_1=l_2=15$,
 (d)  $N=245000$,   $l_1=l_2=15$, (e)  $N=512000$,   $l_1=l_2=16$, and 
 (f)  $N=405000$,   $l_1=l_2=18$.  In all cases the $z$-length of the box trap is $l_3=5$.
These states were calculated in imaginary-time propagation using an analytic  square-lattice initial state with the corresponding number of droplets appropriately arranged on a square lattice. 
 For example, the initial function for a 81-droplet $9\times 9$  square-lattice state is taken  as in  (\ref{ana1}).
In the case of an odd number of droplets,  the size of the unit cell in (a), (d), and (f) are approximately equal, whereas the size of the unit cell in the case of an even number of droplets in (b), (c), and (e) are different and larger than the same in the case of an odd number of droplets.

The formation of a droplet lattice in a   box trap is a bit different from that in a 3D harmonic potential as  used
 in previous experimental \cite{drop1,drop2,2d3,2d2}  and theoretical \cite{2d4,blakieprl,other1,other2,other3}  studies  on  the formation of a droplet lattice. 
 In the case of a harmonically-trapped droplet-lattice state, the number of atoms per droplet is approximately a  constant  depending on the parameters (scattering length and trap frequencies) of the problem;  for $^{164}$Dy atoms with $a=85a_0$, this number is about 11000 atoms/droplet for the trap used in Ref.  \cite{ly}.  This number will be different for a different trap. But in the case of a square box trap  the  number of atoms per droplet can have a wide range of values. It is possible to have a  25-droplet square-lattice state in a square box trap with number of atoms/droplet varying from about 4000 to 9000, provided  the $x$ and $y$ lengths are increased appropriately as the number of atoms is increased.  
If the $x$ and $y$ lengths are not increased, more droplets may appear in place of 25 droplets and the  clean square-lattice structure is destroyed. 
  In Fig. \ref{fig3} we have chosen to show the results for 5000 atoms per droplet in all plots except in Fig. \ref{fig3}(e) with 8000 atoms per droplet.  The possibility of the formation of a droplet lattice state for a small number of atoms in a box trap, as compared to the same in a harmonic trap, could facilitate the use of a box trap in experiments and also in theoretical investigations, as for a  small number of atoms the nonlinearity of the model will be small.

\begin{figure}[t!]
\begin{center}
\includegraphics[width=.49\linewidth]{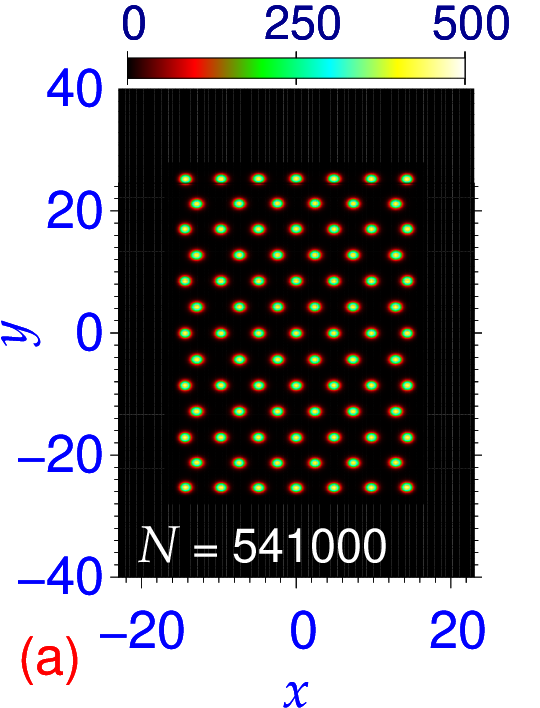}
\includegraphics[width=.49\linewidth]{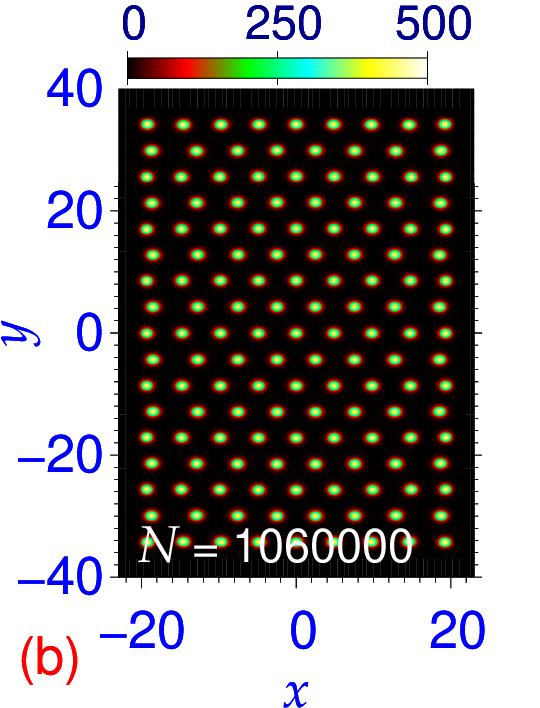}

\caption{ Contour plot of  density of  $^{164}$Dy atoms  $N|\psi(x,y,0)|^2$ in the $x$-$y$ plane  of triangular-lattice crystallization
of (a)  85 droplets  with $x$ length $l_1=16$  and $y$ length $l_2= 27$,   (b)  145 droplets  with $x$ length $l_1=21$  and $y$ length $l_2=36 $, 
 in a 3D rectangular box trap $U_B$. The $z$-length in all cases is $l_3=5$ and $a=85a_0$. }
\label{fig6} 
\end{center} 
\end{figure}

The formation of a supersolid-like state in a square box trap is quite different from that in a harmonic trap; the former gives the unique opportunity to study a spatially-periodic state without the interference of an external trap. This crystallization of matter in free space bounded by a rigid wall, as encountered in a box trap, was the original 
conceptualization of a supersolid \cite{sprsld,sprsld1,sprsld2,sprsld3}. This has the advantage that
by enlarging the boundary of the square box trap it is possible to increase the size of the square-lattice state  as large as we wish without compromising the quality of the lattice and { without the formation of an excessive  atom cloud   as in a harmonic potential trap \cite{ly}.} This is illustrated in Fig. \ref{fig5} for a square box trap $U_A$
for (a) a 169-droplet $13\times 13$ square-lattice state with $N=840000$ atoms and $x$ and $y$ lengths $l_1=l_2= 26$, 
(b) a 196-droplet $14\times 14$ square-lattice state with $N=1180000$ atoms and $x$ and $y$ lengths $l_1=l_2= 28$, and 
(c) a 145-droplet state with $N=1000000$ atoms and $x$ and $y$ lengths $l_1=l_2= 28$. 
In plot (a) [plot(b)], the central site at $x=y=0$ is occupied (unoccupied) and it is a parity-symmetric (parity-antisymmetric) state in $x$ and $y$. 
These states were obtained in imaginary-time simulation using analytic initial functions 
with the same number of droplets as in the final state placed appropriately, viz.  Eq.  (\ref{ana1}).  
The unit cell in (a) and (b) is a square [as in Fig. \ref{fig1}(a)] and in (c) is a centered square [as in Figs. \ref{fig1}(b), and \ref{fig2}(f)].
{
In Fig. \ref{fig5} (c), in-between  nine rows of nine droplets each (present in the initial state), eight rows of eight droplets each are alternately created spontaneously during imaginary-time propagation so as to form a 145-droplet state.} 
In all  cases the droplet-lattice structure is very clean { without an  excessive atom cloud in the background and without 
any visible distortion.  We verified that the reduced atom cloud, we have in a box potential, is of uniform density and which increases the background density by a small amount in all places and do not create any visible atom cloud
in the background. 

Some of these aspects have been studied in Ref. \cite{bt2} for a slightly different box trap, e.g. a box trap in the $x$-$y$ plane and a harmonic trap in the $z$ direction, and for a slightly larger value of the scattering length of $^{164}$Dy atoms, e.g.  $a$ in the range $90a_0$ to  $100a_0$.  Both these differences with   Ref. \cite{bt2} have possibly been fundamental in reducing the non-uniform background atom cloud and obtain very large droplet-lattice structure in this paper. Moreover, a  spontaneous formation of square-lattice state making an angle of 45$^\circ$ with the boundary of the box potential, viz. Figs. \ref{fig2}(f) and 
\ref{fig5}(c),  was not found before. Also, the possibility of the formation of a lattice of different symmetry than that of the binding box potential, viz. Fig. \ref{fig6},  was not also explored  previously.
}

It is intuitively expected that a periodic lattice state with a specific symmetry can efficiently be formed in a box trap with the same symmetry. 
A square box trap is suitable for the formation of a periodic square-lattice state. 
Similarly, a rectangular box trap (not considered here) is also suitable for the formation of a periodic square-lattice state. We will see in the following that a hexagonal box trap is suitable  for the formation of a periodic triangular-lattice state.   Nevertheless, a triangular-lattice state can also be realized in a rectangular box trap with appropriate sizes.  
The inclined square-lattice arrangements of Fig.  \ref{fig2}(f) and \ref{fig5}(c) is made of small triangles with height-to-base ratio 
of  $1:2$.  If  this ratio  could be changed to $\sqrt 3:2$ by squeezing the lattice along $x$  direction, the  square-lattice arrangements of Fig. \ref{fig5}(c) will become a triangular-lattice arrangement.  With this in mind we consider a rectangular box trap    $U_B$   with length to breadth ratio in the $x$-$y$ plane 
$l_1:l_2\approx 1:\sqrt 3$. In such an appropriate trap the 85-droplet square-lattice state becomes a triangular-lattice state as shown in Fig. \ref{fig6}(a) through a
 contour plot of  density of  $^{164}$Dy atoms  $N|\psi(x,y,0)|^2$ in the $x$-$y$ plane 
for $N=541000$ atoms and $x$ and $y$ cut offs $l_1=16$ and $l_2=27$.   In Fig.  \ref{fig6}(b) we display contour plot of density of  a 145-droplet triangular-lattice  state
 for $N=1060000$ atoms and $x$ and $y$ cut offs $l_1=21$ and $l_2=36$. The $z$-length in both cases is  $l_3=5.$
The initial state in imaginary-time simulation was one with the same number of droplets, and with the same symmetry,  as the final state. 
In both these cases, unlike in Fig. \ref{fig5}(c),
  there is some distortion of the
triangular lattice near the boundary, although there is no visible background cloud of atoms.

\begin{figure}[t!]
\begin{center} 
 
\includegraphics[width=\linewidth]{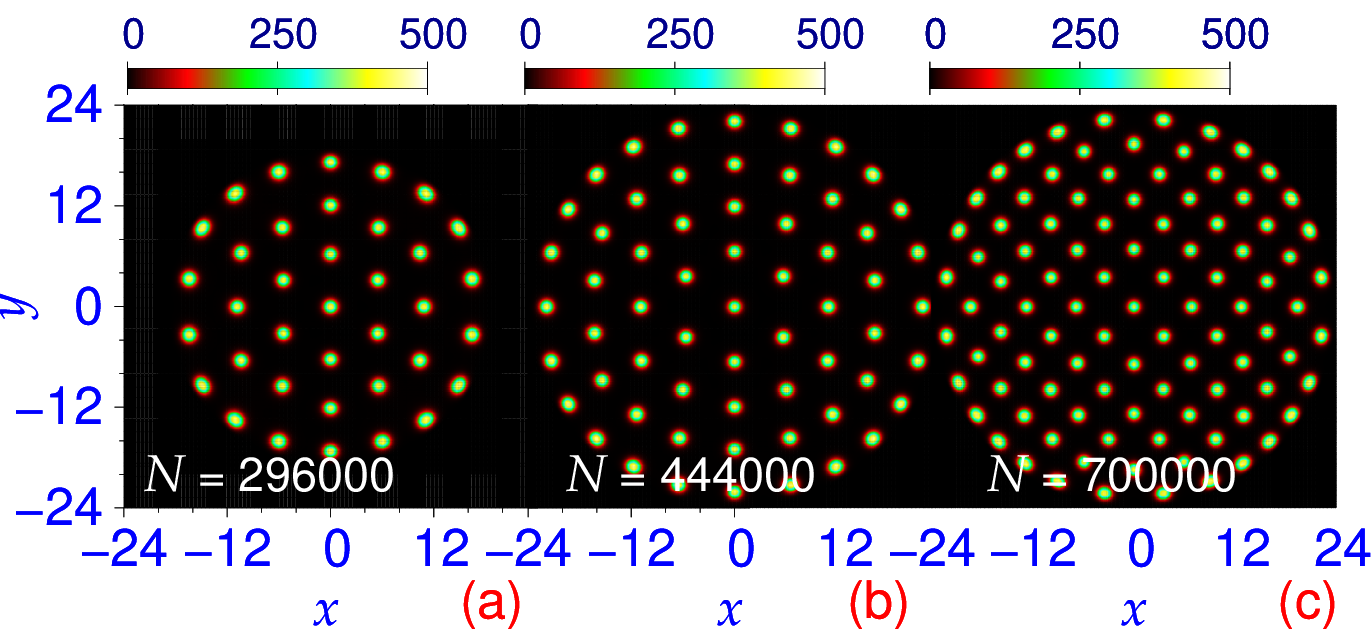}

\caption{ Contour plot of  density of  $^{164}$Dy atoms  $N|\psi(x,y,0)|^2$ in the $x$-$y$ plane  of triangular-lattice crystallization 
of    (a)   37 droplets with $\rho$ length $l_1=19$, and 
(b)  61 droplets with $\rho$ length $l_1=24$,  
and the same of square-lattice crystallization of (c) 97  droplets with $\rho$ length $l_1=24$
 in a 3D cylindrical box trap $U_C$. 
The $z$-length in all cases is $l_3=5$ and $a=85a_0$. }
\label{fig7} 
\end{center}
\end{figure}
 
We also studied the formation of a triangular-  and a square-lattice arrangement of droplets  in a cylindrical box potential
by imaginary-time propagation using an initial state with the same symmetry and with an appropriate number of droplets. The 
 contour plot of  density of  $^{164}$Dy atoms  $N|\psi(x,y,0)|^2$ in the $x$-$y$ plane   of  37- and 61-droplet triangular-lattice states for $N=296000$ and $N=444000$ so obtained  
are shown  in Fig. \ref{fig7}(a)-(b),
for  $\rho\equiv \sqrt{x^2+y^2}$ lengths (a) 19, (b) 24, respectively, with 3, and 4 concentric  hexagonal orbits. The number of droplets in the concentric orbits of Fig. \ref{fig7} are 6, 12, 18, 24, as in the concentric hexagons of a triangular-lattice structure.  
The  contour plot of  density of  $^{164}$Dy atoms  $N|\psi(x,y,0)|^2$ in the $x$-$y$ plane  of a 97-droplet square-lattice state for $N=700000$    
is displayed  in Fig. \ref{fig7}(c),
for  $\rho$ length 24. 
The triangular- and square-lattice arrangements  of the droplets can be clearly  seen in Figs. \ref{fig7}(a)-(b), and (c), respectively,  although there is some distortion of the lattice structure near the circular boundary.

\begin{figure}[t!]
\begin{center}
 
\includegraphics[width=\linewidth]{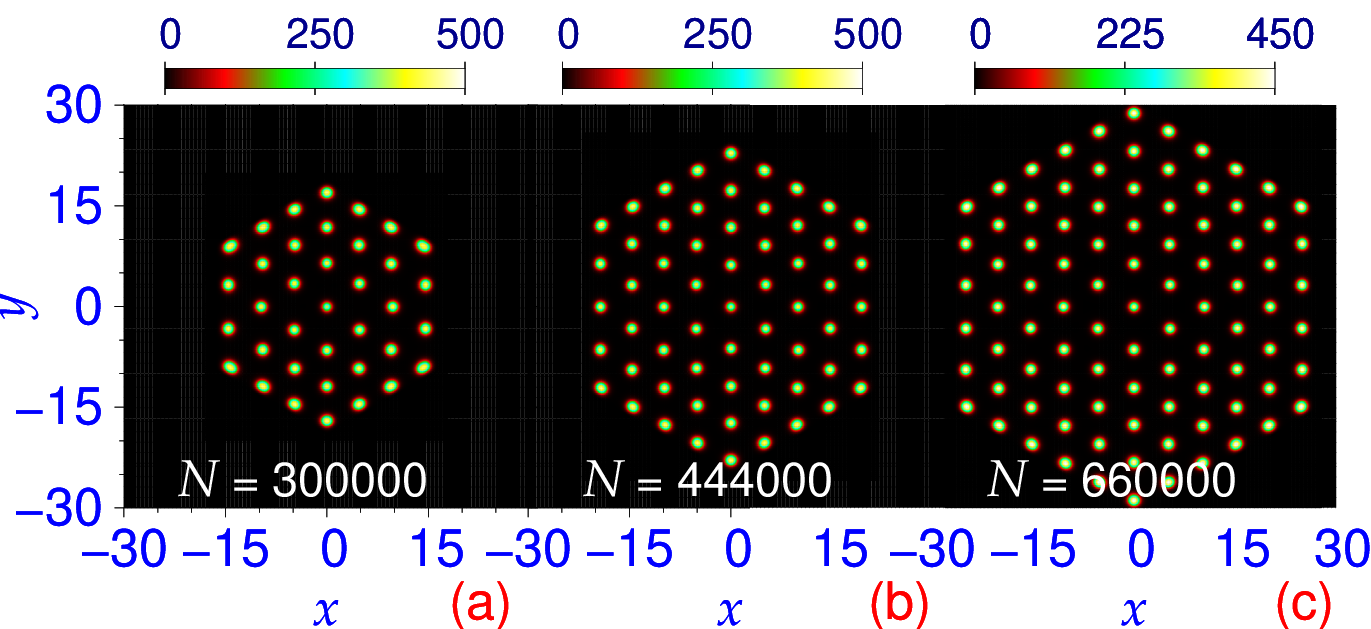}

\caption{Contour plot of  density $|\psi(x,y,0)|^2$ of triangular-lattice crystallization of $^{164}$Dy atoms  in a 3D hexagonal box trap $U_D$  
of (a)  37 droplets  with length  of each side of hexagon $l_1=19$,   (b)   61 droplets with $l_1=25$,
(c)  91 droplets with  $l_1=31 $. The $z$-length in all cases is $l_3=5$ and $a=85a_0$. }
\label{fig8} 
\end{center}
\end{figure}

  The triangular-lattice state in a rectangular or a cylindrical box trap 
 is usually distorted, specially in the outer orbits of the droplets. Next we demonstrate  the formation of perfect triangular-lattice states  in a hexagonal box trap $U_D$ of  appropriate size,   without 
 any distortion. 
 The initial function in imaginary-time simulation was taken to be a triangular-lattice state
with the same number of  droplets as  in the final state. For example, the initial function for a 37-droplet triangular-lattice state is taken  as in Eq. (\ref{ana2}).
 The corresponding contour plot of densities in a hexagonal box trap $U_D$ is illustrated in Fig. \ref{fig8} for (a) 37 droplets with $N=300000$ atoms   and length of each side of hexagon $l_1=19$, (b) 61 droplets with $N=444000$ atoms   and $l_1=25$, and (c) 91 droplets with $N=660000$ atoms   and $l_1=31$. The distortion in the     hexagonal lattice of Fig. \ref{fig8} is  practically invisible compared to that in Figs. (\ref{fig7})(a)-(b) and also in Fig. \ref{fig6}, specially near the boundary of the box traps.

{ To demonstrate the dynamical stability of the droplet-lattice states we perform real-time simulation of the converged imaginary-time state during a long period of time. For this purpose we consider the newly-obtained droplet lattice states displayed in Figs. \ref{fig2}(c) and (f). The triangular-lattice state was already observed in different experiments \cite{drop1,2d2,2d1}, hence it is presumably  dynamically stable.  First we consider the square-lattice state of  Fig.   \ref{fig2}(c)  obtained with a 3D square box trap with $l_1=l_2=18, l_3=5$.  A real-time calculation is performed with the converged imaginary-time state as the initial state  after changing the size of the square box trap to    $l_1=l_2=18.25, l_3=5$.  The square-lattice state expands a bit maintaining the structure of the initial  25-droplet $5\times 5$ state and remains stable for  a long time demonstrating its dynamical stability as illustrated through a contour plot of density 
$N|\psi(x,y,0)|^2$ in Fig. \ref{fig9}(a) at time $t=40$.  Now we consider the 41-droplet state  of Fig.   \ref{fig2}(f)
obtained with a 3D square box trap with $l_1=l_2=18, l_3=5$. To demonstrate its dynamical stability, a real-time calculation is performed with the converged imaginary-time state as the initial state 
 after changing the size of the square box trap to    $l_1=l_2=18.25, l_3=5$.  The  state  remains stable for  a long time demonstrating its dynamical stability as illustrated through a contour plot of density 
$N|\psi(x,y,0)|^2$ in Fig. \ref{fig9}(b) at time $t=40$.

}
\begin{figure}[t!]
\begin{center}
\includegraphics[width=.49\linewidth]{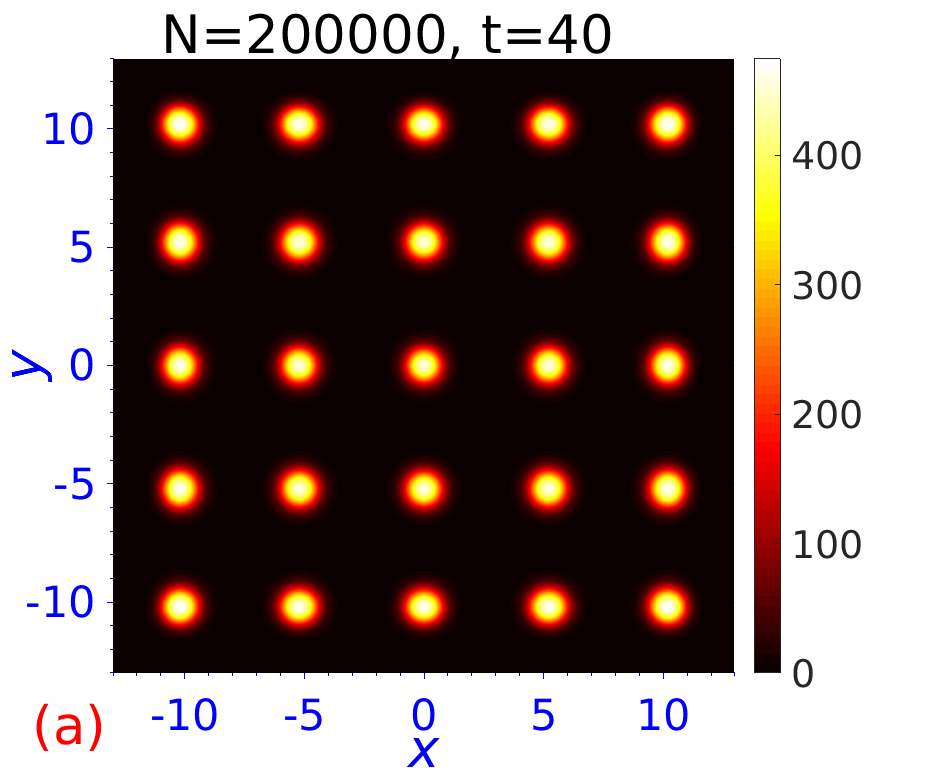}
\includegraphics[width=.49\linewidth]{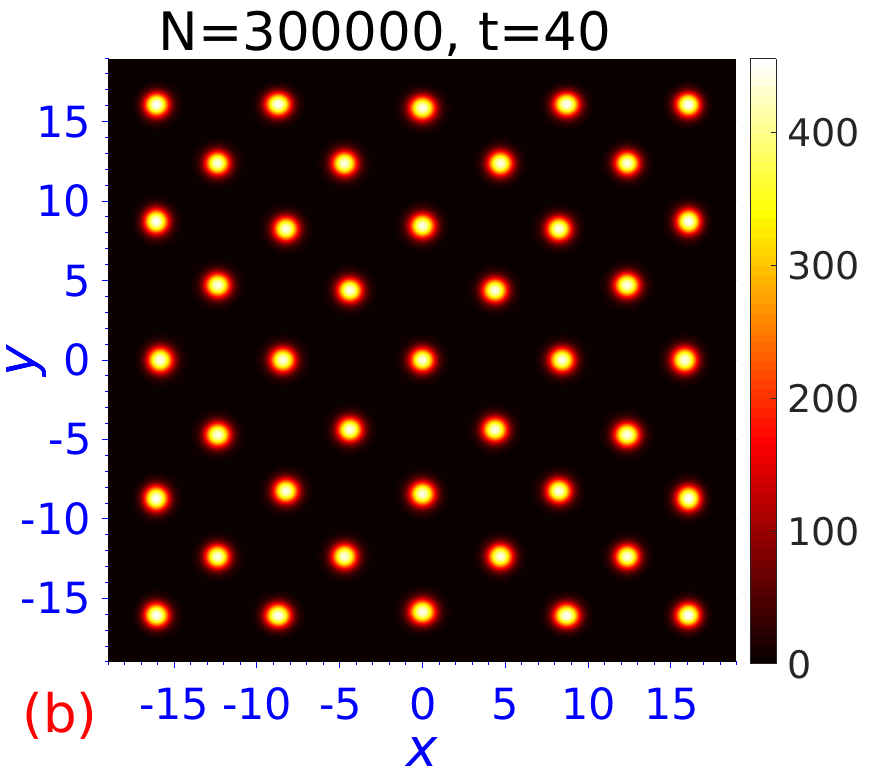}

\caption{ (a)  Contour plot of  density of  $^{164}$Dy atoms  $N|\psi(x,y,0)|^2$ in the $x$-$y$ plane   of the 25-droplet state of Fig. \ref{fig2}(e) in a 3D square  box trap after real-time propagation for $t=40$ time units after changing size of the square box trap from  $l_1=l_2=18, l_3=5$ to  $l_1=l_2=18.25, l_3=5$ at $t=0$. 
 (b)  The same  of the 25-droplet state of Fig. \ref{fig2}(f) in a 3D square  box trap after real-time propagation for $t=40$ time units after changing size of the square box trap from  $l_1=l_2=18, l_3=5$ to  $l_1=l_2=18.25, l_3=5$ at $t=0$. }
\label{fig9} 
\end{center}
\end{figure}

\section{Summary}

\label{IV}

Using  a beyond-mean-field modified GP equation,
we have demonstrated by imaginary-time propagation, including an   LHY-type interaction \cite{qf1,qf2},   supersolid-like spatially-periodic square-lattice and triangular-lattice  crystallization of droplets of a   quasi-2D   dipolar BEC in  square box (cuboid with two equal sides), rectangular box (cuboid), cylindrical box  and hexagonal box (hexagonal  prism) traps. 
The square box and hexagonal box traps are  appropriate for  the formation of  spatially-periodic
square-latice and triangular-lattice  states, respectively, and were used in the numerical investigation of these states. We find two types of square-lattice states in a square box trap, where  
  the square-lattice structure could be aligned parallel to the $x$ and $y$ axes or inclined at
   45$^{\circ}$ $\textdegree$ with these axes.  
Nevertheless, we also studied the formation of triangular-lattice state in a rectangular box and circular box traps as well as the formation of a square-lattice state in a circular box trap. 
The triangular-lattice crystallization in a cylindrical or a rectangular box trap as well as the square-lattice crystallization in a cylindrical box trap  are distorted near the boundary of the box trap. { The dynamic stability of the droplet-lattice states was established by real-time propagation.}

In a harmonically trapped dipolar BEC,  it is difficult to form a very large droplet-lattice state without any deformation \cite{ly} 
whereas in a box trap a large droplet-lattice can be formed without a visible deformation
 and without a perceivable background cloud of atoms. 
Although, the first observation and study of a dipolar supersolid were made in a harmonically-trapped BEC, 
the presence of the interfering external trap possibly leads  to a distorted lattice 
and will thus set a limitation on the formation of a large clean supersolid-like spatially-periodic lattice.   
A box trap, on the other hand, allows the formation of a   spatially-ordered  lattice  of droplets in free space bounded by rigid walls, without any interfering potential, quite similar to an ideal supersolid conjectured in Refs. \cite{sprsld,sprsld1,sprsld2,sprsld3} in free space without any external potential.  A droplet-lattice can be formed in a box potential for a significantly smaller number of atoms compared to the same in a harmonic potential \cite{ly}, which could be an advantage in experimental realization of a large lattice of droplets in a box potential.   
At present moment, square, rectangular, and cylindrical box traps can be realized in a laboratory \cite{16a,16a1,16b,16c,17a,17b,17c,17d} and it is expected that a hexagonal box trap will follow soon.
The results and conclusions of this paper can be tested in experiments with strongly dipolar quasi-2D atomic  BECs of $^{164}$Dy or $^{168}$Er atoms
 with present knowhow.

\section*{Acknowledgments}
SKA acknowledges support by the CNPq (Brazil) grant 301324/2019-0.
LEY-S. would like to
acknowledge the financial support by the Vicerrectoria de Investigaciones - Universidad de Cartagena through Project No.
019-2021.
 
\section*{Data Availability}
 
 Data Availability Statement: No Data associated in the manuscript

%
%



\end{document}